\documentclass{nature}
\usepackage{epsfig}

\def\cr105{2000~CR$_{105}$}
\def\vb12{2003~VB12}

\long\def\symbolfootnote[#1]#2{\begingroup%
\def\thefootnote{\fnsymbol{footnote}}\footnote[#1]{#2}\endgroup} 

\newcommand{\lsun}{{L}_{\odot}}
\newcommand{\msun}{{M}_{\odot}}

\title{An X-ray-emitting blast wave from the recurrent nova RS~Ophiuchi}

\author {J. L. Sokoloski$^1$, G. J. M. Luna$^{1,2}$, K. Mukai$^3$, and
Scott J. Kenyon$^1$}  

\begin{document}

\spacing{1}

\maketitle


\begin{affiliations}
\item Smithsonian Astrophysical Observatory, 60 Garden Street,
Cambridge, MA 02138, USA
\item Instituto de Astronomia, Geof\'\i sica e Ci\^encias Atmosf\'ericas,
Universidade de S\~ao Paulo, Rua do Mat\~ao 1226, Cid. Universit\'aria
05508-900, S\~ao Paulo, Brazil
\item NASA Goddard Space Flight Center and Universities Space Research
Association, Code 662, Greenbelt, MD 20771,
USA
\end{affiliations}

\begin{abstract}
Stellar explosions such as novae and supernovae produce most of the
heavy elements in the Universe.  Although the onset of novae from
runaway thermonuclear fusion reactions on the surface of a white dwarf
in a binary star system is understood\cite{sta78}, the
structure, dynamics, and mass of the ejecta are not well known.  In
rare cases, the white dwarf is embedded in the wind nebula of a
red-giant companion; the explosion products plow through the nebula
and produce X-ray emission.  Early this year, an eruption of the
recurrent nova RS Ophiuchi\cite{ken86,bod87} provided the first
opportunity to perform comprehensive X-ray observations of such an
event and diagnose conditions within the ejecta.  Here we show that
the hard X-ray emission from RS Ophiuchi early in the eruption
emanates from behind a blast wave, or outward-moving shock wave, that
expanded freely for less than 2 days and then decelerated due to
interaction with the nebula.  The X-rays faded rapidly, suggesting
that the blast wave deviates from the standard spherical shell
structure\cite{sed59,che82,bod85}.  The early onset of deceleration
indicates that the ejected shell had a low mass, the white dwarf has a
high mass\cite{yar05}, and that RS Ophiuchi is a progenitor of the
type of supernova integral to studies of the expansion of the
universe.
\end{abstract}

On February 12, 2006, RS Ophiuchi (RS~Oph) was discovered in outburst
near the historical maximum optical brightness of 4.4
magnitudes\cite{nar06} -- more than 4 times brighter than the
threshold for naked-eye visibility.  The Rossi X-Ray Timing Explorer
(RXTE) satellite\cite{jah96} first detected strong hard X-rays from
RS~Oph -- the strongest ever observed from a white dwarf -- at peak
X-ray brightness on outburst day 2 with the All Sky Monitor camera.
RXTE subsequently measured the X-ray spectrum with the Proportional
Counter Array (PCA) on outburst days 3, 6, 10, 14, 17, and 21
(Fig.~1).  During this time, the binary was in the phase of its orbit
in which the white dwarf and red giant appear side-by-side from the
perspective of the earth\cite{dob94,fek00}.

During the first three weeks of the outburst, the total X-ray
luminosity from the hottest X-ray emitting gas fell from $\sim 300\;
\lsun$ to $\sim 30\; \lsun$ (integrated from 0.5 to 20 keV, and taking
a distance of 1.6 kpc\cite{hje86,sni87}).  From day 6 onward, the
X-ray emission from the hottest gas decreased as $t^{-5/3}$, where $t$
is the time since the beginning of the outburst (Fig.~2a).  After
rising to maximum in about 2 days, the emission measure ($n^2
\times V$, where $n$ is the density of radiating electrons and $V$ is the
X-ray emitting volume) remained roughly constant at $\sim 10^{58}$
cm$^{-3}$ until day 6, after which time it fell as $t^{-4/3}$.  For
comparison, the X-ray luminosity from hot gas in classical novae,
where the white dwarf is not embedded in a dense nebula, is always
below $30\; \lsun$\cite{muk01,gre03}, and the classical nova with the
most comprehensive X-ray observations\cite{ori04} (V1974 Cyg) took
more than 100 days to reach a maximum X-ray emission measure of a few
times $10^{56}\;{\rm cm}^{-3}$.  As the outburst progressed and the
X-ray luminosity decreased, the bulk of the X-ray emission from RS~Oph
shifted to lower energies (see Fig.~1).  This shift corresponds to a
drop in the post-shock temperature.  From the time of the first
spectral observation on day 3, this temperature decreased as
$t^{-2/3}$ (Fig.~2b).

If the X-rays are from circumstellar material heated by the blast wave
produced in the nova explosion, we can associate the characteristic
X-ray temperature with the blast wave speed via the standard
strong-shock relation.  Optical emission-line widths directly measure
the motion of either gas behind the blast-wave shock that has
undergone charge exchange\cite{che80} or the ejecta, and these line
widths\cite{iij06} support the shock speeds we derive.  Since $v
\propto T^{1/2}$, where $T$ is the post-shock plasma temperature and
$v$ is the blast-wave speed, the rate of temperature decrease
indicates that the blast-wave speed decreased as $t^{-1/3}$.  This
direct measurement of the shock deceleration rate is the first for any
nova.  For an initial shock speed of 3,500
km/s\cite{bui06,fuj06},
and extrapolating backwards from our inferred
shock speeds, we conclude that the deceleration began at approximately
day 1.7.  With the measured deceleration, we can obtain the blast wave
radius as a function of time.  Comparing the blast wave radius on days
21 and 27 to radio images of the expanding shock on those days (Rupen,
Mioduszewski, and J.L.S., in preparation), we confirm the 1.6 kpc
distance to RS~Oph.

The standard blast wave theory presupposes that the shock structure is
self-similar, or that the density and temperature as a function of
fractional distance from the white dwarf to the shock front are
similar to these functions at later times.  In the initial
ejecta-dominated phase, the ejecta expand freely and produce a shock
moving at constant velocity.  After the expanding blast wave has swept
up a few times the ejecta mass, it begins to decelerate and enters the
Sedov-Taylor (ST) phase.  After the 1985 outburst of RS~Oph, standard
supernova blast wave theory was adapted to explain the late-time X-ray
emission\cite{bod85}.  Models for RS~Oph must also account for the
fact that the nova explosion takes place within a stellar wind whose
density far from the binary decreases as the inverse of the distance
from the binary squared ($n \propto 1/r^2$, where $r$ is the distance
from the wind-producing red giant).  Whereas the early hard X-ray
emission from classical novae is thought to arise in the ejecta that
has been heated by a reverse shock\cite{obr94}, the early hard X-ray
emission from RS~Oph comes predominantly from circumstellar material
heated by the forward-moving blast wave.

Our observations indicate that the standard self-similar blast-wave
model does not explain the ejection of material from this nova.
Although our discovery that $v \propto t^{-1/3}$ agrees with the
theoretical prediction for a blast wave moving into a stellar wind and
the general expansion law for a shock wave resulting from a point
explosion\cite{sed59,che82,bod85}, the observed rate of X-ray fading
disagrees with predictions.  In the current outburst, the initial
X-ray flux was almost 100 times brighter than expected based on
observations and modeling of the late-time X-ray emission in the 1985
outburst\cite{bod85,mas87}.  The total X-ray flux for thermal
bremsstrahlung emission from a fully ionized gas is proportional to
$T^{1/2} n^2 V$.  If the volume of the post-shock emitting region
increases as the radius of the blast wave cubed (as in the standard
similarity solutions for supernova shells), the X-rays should decrease
as $t^{-1}$.  The X-ray emission from RS~Oph in the ST phase decayed
instead as $t^{-5/3}$.

The X-ray flux evolution reflects the environment of the binary and
suggests how the nova explosion deviates from a spherically symmetric,
self-similar blast wave.  The fast X-ray rise of RS~Oph compared to
classical novae is due to the higher circumstellar density and the
faster initial shock speed.  The observed peak emission measure can be
attained in 2 days if the material swept up by the blast wave produces
thermal bremsstrahlung emission at the immediate post-shock density
and temperature.  The observed X-ray decay rate from day 6 onward is
consistent with emission from a hot post-shock region whose volume
increases roughly as the square of the blast-wave radius.  Such an
emitting-volume evolution could be produced by radiative cooling,
which has little impact on the early-time shock velocity, but a
significant impact on the post-shock temperature
structure\cite{obr87}.  Alternatively, or in conjunction with
radiative cooling, a non-spherical ejecta geometry such as that
inferred from radio observations\cite{tay89,llo93}, or a contribution
to the X-ray luminosity from ejecta heated by a reverse shock, could
also steepen the X-ray decline.

Finally, since the transition to the ST phase occurs when a few times
the ejecta mass have been swept up, the timing of the onset of
deceleration constrains the ejected shell mass independently of all
previous estimates.  Taking a density inside the binary of $10^9\;
{\rm cm}^{-3}$ from observational constraints on the red-giant
mass-loss rate\cite{dob94} and assuming that the density fall-off does
not begin until outside the binary, on $\sim$day 2 the shock had swept
up only a few times $10^{-7}\;
\msun$ of material.  Thus, the ejecta mass could not have been much
more than $10^{-7}\; \msun$.  Theoretical models of nova explosions
with recurrence times of ~20 years, as in RS~Oph, have been
constructed for white dwarfs with masses of 1.25 and
1.40~$\msun$\cite{yar05}.  Whereas the 1.25~$\msun$ models require an
ejecta mass of $\sim 10^{-6}\; \msun$, the 1.40~$\msun$ models eject
material with a total mass of $2
\times 10^{-7}\; \msun$.  The ejected shell mass of $\sim 10^{-7}\; \msun$
from RS~Oph is thus more consistent with the 1.4~$\msun$ model.

The mass accumulation efficiency of white dwarfs in recurrent novae is
controversial\cite{yar05,hac01}.  Our conclusion that the white dwarf
in RS~Oph must be extremely close to the maximum mass shows that
significant mass accumulation is possible, and that some recurrent
novae will explode as Type Ia supernovae.  If recurrent novae lead to
supernovae Ia such as the one recently discovered to contain hydrogen
in its spectrum\cite{ham03}, sweeping up of the red-giant wind by the
nova blast wave could explain the cavity surrounding this
supernova\cite{woo04,woo06}.  Our results also have bearing on studies
of Type II supernovae.  The early days of the 2006 outburst of RS~Oph
show the first two phases of blast wave evolution, the analog of which
can take hundreds of years for a supernova remnant.  They also provide
the blast-wave deceleration law that is difficult to measure directly
in supernovae.  Our results should motivate modeling that will lead to
better understanding of cooling mechanisms as well as post-shock and
ejecta structure of novae and other stellar explosions.

\clearpage

\begin{figure}
\vspace*{-3cm}
\epsfxsize 5.0in
\begin{center}
\epsffile{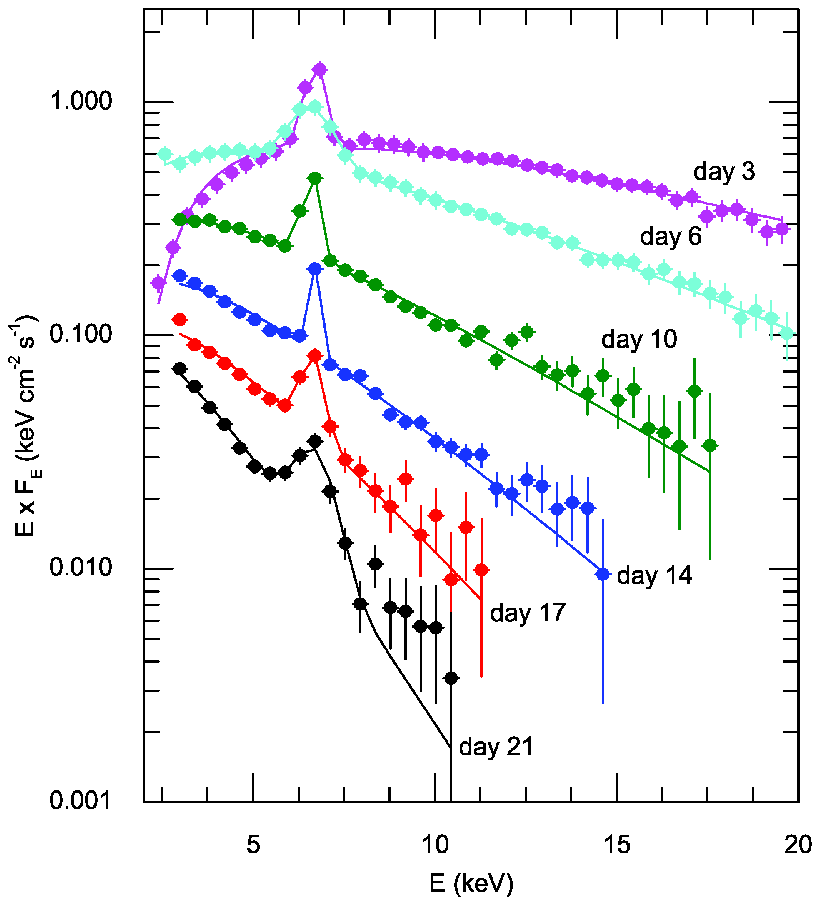}
\end{center}
\end{figure}

\vspace{-1.5cm}
\noindent
{\bf Figure 1} X-ray spectra from the first 3 weeks of the 2006
outburst of RS~Oph.  These six spectra were taken with the PCA
instrument on board the RXTE satellite.  The abscissa is the energy,
$E$, of the detected X-rays.  The quantity $E\times F_E$ is the photon
energy times the energy flux density, and it provides a measure of the
relative amount of energy being emitted across the spectrum.  The
usable energy range for the PCA is approximately 2-25 keV.  The
spectra were reduced using standard data reduction software (the
FTOOLS package) and the standard RXTE background models.  The presence
of a blended emission line from H-like and He-like Fe indicates that
the X-ray emission is optically thin thermal plasma emission; the
spectra can all be reasonably well fit (reduced $\chi^2$ ranging from
0.4 to 1.7 for the 6 observations) with a single-temperature thermal
bremsstrahlung model plus line emission from Fe and absorption by
intervening material. Although RXTE is not very sensitive to
absorption, which preferentially affects photons with energies less
than around 2~keV, the absorption was high enough on day 3 ($N_H =
[5.5 \pm 1.1]
\times 10^{22}\; {\rm cm}^{-2}$, where $N_H$ is the column density of
neutral hydrogen) to have a significant impact on the low-energy end
of the spectrum.  The absorption lessened in the subsequent
observations as the blast wave moved outward and the amount of neutral
intervening wind decreased.  After day 10, we fixed the absorption
parameter in our fits to the interstellar
value\cite{mas87,dav87,sni87} of $N_H = 4 \times 10^{21}\; {\rm
cm}^{-2}$.  Error bars represent 1 standard deviation.

\begin{figure}
\epsfxsize 6.5in
\begin{center}
\epsffile{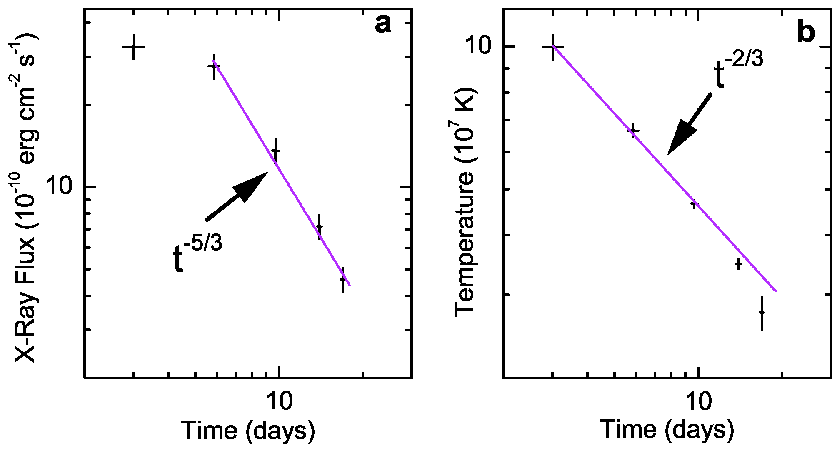}
\end{center}
\end{figure}

\noindent
{\bf Figure 2} X-ray flux and post-shock plasma temperature as a
function of time.  We measure the time, $t$, from the start of the
outburst, which we take to be February 12.6 U.T.  {\bf a}, X-ray flux
in the energy range 0.5-20 keV from the hottest plasma, obtained by
integrating the best-fit thermal bremsstrahlung model, corrected for
the effects of absorption by intervening material.  This energy range
was chosen to match typical observing ranges and to include most of
the emission from the hot plasma.  The uncertainty was taken to be
5\%, and comes primarily from uncertainty in the model parameters,
especially interstellar absorption.  {\bf b}, The characteristic X-ray
plasma temperature determined from fitting the X-ray spectra with
thermal bremsstrahlung models, as a function of time.  Error bars
represent 1 standard deviation.  The temperature decays as expected
for material that has been shock-heated by a decelerating Sedov-Taylor
blast wave moving into a $n \propto 1/r^2$ stellar wind.  Since RXTE
is sensitive to X-rays with energies greater than 2 keV, we obtain the
best determination of the parameters from the first several
observations, when the post-shock plasma is hottest.  By the sixth
observation, the error bars have become too large for these data to be
useful (see Fig.~1), so the data from the final observation are not
included on the plots.

\begin{addendum}
 \item We are grateful for discussions with J. Raymond, C. Rakowski,
 M. Wood-Vasey, and C. Matzner.  J.L.S. acknowledges support from the
 NSF.
G.J.M.L. acknowledges support from NASA, plus CNPq and FAPESP (Brazil).
 \item[Competing Interests] The authors declare that they have no
competing financial interests.
 \item[Correspondence] Correspondence and requests for materials
should be addressed to \\
J.L.S. (jsokoloski@cfa.harvard.edu).
\end{addendum}

\clearpage

\end{document}